 \documentclass[aps,twocolumn,amsmath,amssymb,groupedaddress,float]{revtex4}

\usepackage{graphicx} 

\begin{document}

\title{Quantum Monte Carlo calculation of the equation of state of
neutron matter}

\author{S.~Gandolfi}
\email{gandolfi@sissa.it}
\affiliation{International School for Advanced Studies, SISSA
Via Beirut 2/4 I-34014 Trieste, Italy}
\affiliation{INFN, Sezione di Trieste, Trieste, Italy}
\author{A.~Yu.~Illarionov}
\email{illario@sissa.it}
\affiliation{International School for Advanced Studies, SISSA
Via Beirut 2/4 I-34014 Trieste, Italy}
\affiliation{INFN, Sezione di Trieste, Trieste, Italy}
\author{K.~E.~Schmidt}
\email{kevin.schmidt@asu.edu}
\affiliation{Department of Physics, Arizona State University, Tempe, AZ 85287,
USA}
\author{F.~Pederiva}
\email{pederiva@science.unitn.it}
\affiliation{Dipartimento di Fisica dell'Universit\'{a} di Trento,
via Sommarive 14, I--38050 Povo, Trento, Italy}
\affiliation{INFN, Gruppo Collegato di Trento, Trento, Italy}
\author{S.~Fantoni}
\email{fantoni@sissa.it}
\affiliation{International School for Advanced Studies, SISSA
Via Beirut 2/4 I-34014 Trieste, Italy}
\affiliation{INFN, Sezione di Trieste, Trieste, Italy}
\affiliation{INFM {\sl DEMOCRITOS} National Simulation Center, Via Beirut 2/4
I-34014 Trieste,  Italy}

\begin{abstract}
We calculate the equation of state of neutron matter at zero
temperature by means of the auxiliary field diffusion Monte Carlo method
(AFDMC) combined with a fixed-phase approximation.  The calculation of
the energy is carried out by simulating up to 114 neutrons in a periodic
box.  Special attention was made to reduce finite size effects at the
energy evaluation by adding to the interaction the effect due to the
truncation of the simulation box, and by performing several simulations
using different number of neutrons.  The finite size effects due to
the kinetic energy were also checked by employing the twist--averaged
boundary conditions.  We considered a realistic nuclear Hamiltonian
containing modern two-- and three--body interactions of the Argonne
and Urbana family.  The equation of state
can be used to compare and to calibrate
other many-body calculations and to predict properties of neutron stars.
\end{abstract}

\date{\today}

\maketitle

\section{Introduction}
The equation of state of nuclear matter and its properties plays
a central role in the modeling of neutron stars\cite{raffelt96}. The
density in the star ranges from a small fraction of, up to several times,
nuclear saturation density, 0.16 fm$^{-3}$, which is found in the
center of heavy nuclei.  At such extreme conditions no phenomenological
data determined from experiments are available, and because the matter
inside the neutron stars is closer to neutron matter than symmetric
nuclear matter, heavy-ion collision experiments do not
substantially 
constrain the equation of state\cite{danielewicz02}.
A realistic calculation of the equation of state
of neutron matter is then particularly challenging in both many-body
nuclear physics and astrophysics.

The equation of state
of neutron matter can in principle be computed in the framework
of many-body theories using a bare interaction. A common alternative
is represented by effective Skyrme forces.  However, the resulting
equation of state
strongly depends on the parameters of the effective interaction used,
even in the low density regime\cite{brown00}.  At present, there are a
wide range and type of Skyrme interactions.  However their non-realistic
character impairs their ability to reliably calculate the properties
of neutron stars\cite{stone03}. More accurate many-body techniques are
then needed to perform predictive calculations.

A microscopic calculation of neutron matter starting from a
non-relativistic nucleon-nucleon and three-nucleon interaction
is both challenging and of great relevance.  Variational techniques based
on correlated basis functions are good candidates to solve
for the ground-state of neutron matter. The operatorial structure of
the nuclear Hamiltonian and the strong correlations induced by the high
density make these techniques hard to use.  The energy evaluation using
the correlated basis function
theory is usually performed by solving the Fermi Hyper Netted
Chain (FHNC) equations\cite{pandharipande79} neglecting many elementary
diagrams.  In addition, the operatorial structure of the Hamiltonian
leads to additional approximations, like the Single Operator Chain (SOC)
approximation, due to the non-commutativity of the terms entering in
the variational wave function.  Therefore the resulting equation of state
contains, in
principle, uncontrolled approximations which may be partially corrected
by computing the energy exactly up to a few first orders in the cluster
expansion \cite{morales02}.

Despite the progress of the last several years in the determination of
sophisticated two- and three-nucleon interactions,
large discrepancies among different
calculations of nuclear and neutron matter are still present.  Quantum
Monte Carlo techniques based on projection can be very accurate for
calculating the ground state and low lying excited states of nuclei. In
particular Green's Function Monte Carlo (GFMC) was employed to fit the
three-nucleon interaction
form in nuclei up to $A=8$\cite{pieper01}, and then used to test
the nuclear Hamiltonian up to the $^{12}$C ground-state\cite{pieper05}.
At present the huge number of required numeric operations limits the
applicability of this method to only about 14 neutrons\cite{carlson03}.

The Auxiliary Field Diffusion Monte Carlo (AFDMC\cite{schmidt99}) combined
with a fixed-phase approximation was employed to predict properties
of nuclei in very good agreement with the GFMC\cite{gandolfi07b},
and stressed important limitations of other many-body theories used in
nuclear matter calculations\cite{gandolfi07}.

In this work we present an accurate evaluation of the equation of state
of neutron
matter using a realistic two- and three-nucleon interactions
(the Argonne $v_8^\prime$
and Argonne $v_{18}$ combined with Urbana-IX\cite{pudliner97}).
The computed equation of state
can be used as
a benchmark for other many-body techniques.

The plan of the paper is the following: in the next section we describe
the structure of the nuclear Hamiltonian we used; in Sec. \ref{sec:method}
we will briefly review the AFDMC method and explain the fixed-phase
approximation. The results will be presented in Sec. \ref{sec:results}
and some conclusions will be given in the last section.

\section{Hamiltonian}
Properties of a generic nuclear system can be studied starting from the
non-relativistic Hamiltonian
\begin{equation}
\label{eq:hamiltonian}
H=-\frac{\hbar^2}{2m}\sum_i\nabla_i^2+\sum_{i<j}v_{ij}+\sum_{i<j<k}V_{ijk} \,,
\end{equation}
which includes the kinetic energy operator, a two-nucleon interaction
$v_{ij}$, and a three-nucleon interaction $V_{ijk}$.

The nucleon-nucleon interactions are usually dependent on the relative spin and
isospin state of the nucleons and therefore written as a sum of several
operators.  The coefficients and radial functions that multiply each
operator are adjusted by fitting experimental scattering data, and
the type and number of these operators depends on the interaction.
A large amount of empirical information about the nucleon-nucleon
scattering problem has been accumulated. In 1993, the Nijmegen group
analyzed all nucleon-nucleon scattering data below 350 MeV published in physics
journals between 1955 and 1992\cite{stoks93b}.  Nucleon-nucleon
interaction models
that fit the Nijmegen database with a $\chi^2/N_{data}\sim$1 are called
``modern'' which include the Nijmegen models\cite{stoks94} (Nijm93, Nijm
I, Nijm II and Reid-93), the Argonne models\cite{wiringa95,wiringa02}
and the CD-Bonn\cite{machleidt96}.  However all of these interactions,
when used alone, underestimate the triton binding energy, suggesting that
the contribution of a three-nucleon force is essential to reproduce the
physics of nuclei.

The most sophisticated Argonne interaction is the
Argonne $v_{18}$\cite{wiringa95}
potential, written as a sum of 18 operators.  However we often consider
another interaction, the Argonne $v_8^\prime$\cite{wiringa02} that is a
simplified version of Argonne $v_{18}$;
it contains only 8 operators, and the prime
symbol indicates that such potential is not just a simple truncation of
Argonne $v_{18}$,
but a reprojection, which preserves the isoscalar part
in all S and P partial waves as well as in the $^3D_1$ wave and its
coupling to $^3S_1$.  The Argonne $v_8^\prime$ is a bit more attractive than
Argonne $v_{18}$
in light nuclei by about 0.5 MeV per nucleon\cite{pudliner97},
but its contribution is very similar to Argonne $v_{18}$ in neutron drops,
where the difference is about 0.06 MeV per neutron\cite{pieper01}.

The Argonne potential between two nucleons $i$ and $j$ is written in
the coordinate space as a sum of operators
\begin{equation}
\label{eq:vop}
v_{ij}=\sum_{p=1}^n v_p(r_{ij})O_{ij}^p \,,
\end{equation}
where $n$ is the number of operators which depends on the potential, $v_p(r)$
are radial functions, and $r_{ij}$ is the inter-particle distance.

The eight operators included in Argonne $v_8^\prime$ give the largest
contributions to the nucleon-nucleon
interaction. The first six of them come from
the one-pion exchange between nucleons, while the last two terms depend
on the velocity of nucleons and give the spin-orbit contribution.
These eight operators are
\begin{equation}
O_{ij}^{p=1,8}=(1,\boldsymbol\sigma_i\cdot
\boldsymbol\sigma_j,S_{ij},\boldsymbol L_{ij}\cdot\boldsymbol S_{ij})
\times(1,\boldsymbol\tau_i\cdot\boldsymbol\tau_j) \,,
\end{equation}
where $S_{ij}$ is the tensor operator
\begin{equation}
S_{ij}=3(\boldsymbol\sigma_i\cdot\hat r_{ij})
(\boldsymbol\sigma_j\cdot\hat r_{ij})
-\boldsymbol\sigma_i\cdot\boldsymbol\sigma_j \,,
\end{equation}
$\boldsymbol L_{ij}$ is the relative angular momentum of couple $ij$
\begin{equation}
\boldsymbol L_{ij}=\frac{1}{2i}(\boldsymbol r_i-\boldsymbol r_j)
\times(\boldsymbol\nabla_i-\boldsymbol\nabla_j) \,,
\end{equation}
and $\boldsymbol S_{ij}$ is the total spin of the pair
\begin{equation}
\boldsymbol S_{ij}=\frac{1}{2}(\boldsymbol\sigma_i+\boldsymbol\sigma_j) \,,
\end{equation}
with
both $\boldsymbol L_{ij}$ and $\boldsymbol S_{ij}$ divided by $\hbar$
to make them unitless.

In modern interactions these eight operators are the standard ones
required to fit S and P wave scattering data in both triplet and singlet
isospin states.

The three-nucleon interaction
contribution is mainly attributed to the possible $\Delta$
intermediate states that an excited nucleon could assume after and before
exchanging a pion with other nucleons.  This process can be written as
an effective three-nucleon interaction, and its parameters are fit to
light nuclei\cite{carlson81,pieper01} and eventually to properties of
nuclear matter, such as the empirical equilibrium density and the energy
at saturation\cite{carlson83}. The three-nucleon interaction
must accompany the two-nucleon interaction
and the total Hamiltonian studied.

The Urbana IX three nucleon interaction used in our calculation has the
following form
\begin{equation}
V_{ijk}=V_{2\pi}+V_R \,.
\end{equation}
The Fujita-Miyazawa term\cite{fujita57} is spin-isospin dependent:
\begin{eqnarray}
\label{eq.fm}
V_{2\pi}=A_{2\pi}\sum_{cyc}\Big[\{X_{ij},X_{jk}
\{\tau_i\cdot\tau_j,\tau_j\cdot\tau_k\}+
\nonumber \\
\frac{1}{4}[X_{ij},X_{jk}][\tau_i\cdot\tau_j,\tau_j\cdot\tau_k]\Big] \,,
\end{eqnarray}
where
\begin{align}
X_{ij}&=Y(m_\pi r_{ij})\boldsymbol\sigma_i\cdot\boldsymbol \sigma_j
+T(m_\pi r_{ij})S_{ij} \,,
\nonumber \\
Y(x)&=\frac{e^{-x}}{x}\xi_Y(r) \,,
\nonumber \\
T(x)&=\left(1+\frac{3}{x}+\frac{3}{x^2}\right)Y(x)\xi_T(r) \,,
\nonumber \\
\xi_Y(r)&=\xi_T(r)=1-e^{-cr^2} \,.
\end{align}
The phenomenological $V_R$ part is
\begin{equation}
V_{ijk}^R=U_0 \sum_{cyc}T^2(m_\pi r_{ij})T^2(m_\pi r_{jk}) \,.
\end{equation}
For neutrons, the commutator terms in Eq. \ref{eq.fm} are zero, and
each of the anticommutator terms has only spin operators for two of the
three neutrons.

The $A_{2\pi}$ term of Urbana-IX was originally fitted, along with the
Argonne $v_{18}$ parameters,
to reproduce the triton and alpha particle binding energy, while the $U_0$
strength was adjusted to obtain the empirical equilibrium density of
nuclear matter\cite{pudliner95}. However, while the ground state of light
nuclei can be exactly solved with few-body techniques, the determination
of the equation of state of symmetric nuclear matter can be evaluated
only using many-body techniques that contain uncontrolled approximations.

\section{Method}
\label{sec:method}
\subsection{Diffusion Monte Carlo}
The Auxiliary Field Diffusion Monte Carlo method is an extension of
the usual Diffusion Monte Carlo to deal with Hamiltonians that
are spin-isospin dependent.  The diffusion Monte Carlo
method\cite{guardiola98,mitas99},
projects out the ground state properties by starting from a trial wave
function not orthogonal to the true ground state.

Consider a generic trial wave function $\psi_T$ expanded over a
set $\{\phi_n\}$ of eigenstates of the Hamiltonian:
\begin{equation}
\psi_T(\bold R)=\psi(R,0)=\sum_n c_n \phi_n(\bold R) \,.
\end{equation}
The propagation in imaginary time $\tau$ is given by
\begin{equation}
\label{eq:a}
\psi(\bold R,\tau)=\sum_n c_n e^{-(H-E_0)\tau}\phi_n(\bold R) \,,
\end{equation}
where $E_0$ is a normalization factor, and $\bold R$ represent the
spatial coordinates of the system.  In the limit $\tau\rightarrow\infty$,
$\psi(\bold R,\tau)$ approaches the lowest eigenstate $\phi_0$ with
the same symmetry as $\psi$.  The evolution can be done by solving the
integral equation
\begin{equation}
\label{eq:intschr}
\psi(\bold R,\tau)=\int G(\bold R,\bold R',\tau)\psi(\bold R',0)d\bold R' \,,
\end{equation}
where the wave function is described with a set of $N_w$ configurations
called \emph{walkers} as following:
\begin{equation}
\langle \bold R|\psi\rangle =
\psi(\bold R) \cong
\sum_{k=1}^{N_w} \langle \bold R|\bold R_k\rangle \langle \bold R_k|\psi\rangle
\end{equation}
and
\begin{equation}
\langle \bold R|\bold R_k\rangle=\delta(\bold R-\bold R_k) \,.
\end{equation}
The kernel $G(\bold R,\bold R',\tau)$ is the Green's function of the
system, and can be expressed as the matrix element
\begin{equation}
G(\bold R,\bold R',\tau)=
\langle \bold R\lvert e^{-(H-E_0)\tau}\rvert \bold R'\rangle \,.
\end{equation}
By considering a generic Hamiltonian and the Trotter decomposition,
the form of the Green's function in the small imaginary time-step limit
$\Delta\tau \rightarrow 0$ is
\begin{equation}
G(\bold R,\bold R',\Delta\tau)
\approx e^{-\frac{V(\bold R)+V(\bold R')}{2}\Delta\tau}
G_0(\bold R,\bold R',\Delta\tau) \,,
\end{equation}
where $G_0$ is the Green's function of the noninteracting system
\begin{equation}
G_0(\bold R,\bold R',\tau)=\left(\frac{m}{2\pi\hbar^2\tau}\right)^\frac{3A}{2}
e^{-\frac{m|\bold R-\bold R'|^2}{2\hbar^2\tau}} \,,
\end{equation}
and the factor due to the interaction plus the trial eigenvalue $E_T$
is the normalization of the Green's function factor computed over the
time interval $\tau$:
\begin{equation}
\label{eq:weight}
w=e^{-\left(\frac{V(\bold R)+V(\bold R')}{2}-E_T\right)\Delta\tau} \,.
\end{equation}

The integral equation \ref{eq:intschr} can be solved in a Monte Carlo way.
At each time-step all walkers are moved with the diffusion term of the
free Green's function $G_0$, so that for each walker a new set $\bold R'$
of spatial coordinates are generated according to
\begin{equation}
\bold R'=\bold R+\boldsymbol\eta \,,
\end{equation}
where $\bold R$ is the old configuration, and $\boldsymbol\eta$ is a
vector of random numbers with probability density $G_0$.

The normalization of Eq. \ref{eq:weight}, translated into a weight of
the walker, is sampled using the \emph{branching} technique in which $w$
gives the probability of a configuration to multiply at the next step.
Computationally, this is implemented by weighting estimators according
to $w$, and generating from each single walker a number of replicas
\begin{equation}
n=\left [w+\xi \right ] \,,
\end{equation}
where $\xi\in[0;1]$ is a random number and $[x]$ means integer part of $x$.

The infinite imaginary-time limit is reached by iterating this process
for a sufficient total time $\tau=n\Delta\tau$.

\subsection{Auxiliary Field Diffusion Monte Carlo}
\label{subsec:AFDMC}

In the case of nuclear Hamiltonians the potential contains quadratic
spin and isospin and tensor operators, so the many-body wave function
cannot be written as a product of single particle spin-isospin states.

For instance, let us consider the generic quadratic spin
operator $\boldsymbol\sigma_i\cdot\boldsymbol\sigma_j$ where the
$\boldsymbol\sigma$ are Pauli's matrices operating on particles.  It is
possible to write
\begin{equation}
\boldsymbol\sigma_i\cdot\boldsymbol\sigma_j=2P_{ij}^\sigma-1 \,,
\end{equation}
where $P_{ij}^\sigma$ interchanges two spins, and this means that the wave
function of each spin-pair must contain both components in the triplet
and singlet spin-state\cite{pieper98,carlson99b}.  By considering all
possible nucleon pairs in the systems, the number of possible spin-states
grows exponentially with the number of nucleons.

Thus, in order to perform a diffusion Monte Carlo
calculation with standard nuclear
Hamiltonians, it is necessary to sum over all the possible single particle
spin-isospin states of the system to build the trial wave function used
for propagation.  This is the standard approach in GFMC calculations
for nuclear systems.

The idea of AFDMC is to rewrite the Green's function in order to change
the quadratic dependence on spin and isospin operators to a linear
dependence by using the Hubbard-Stratonovich transformation.

For neutrons $\boldsymbol\tau_i\cdot\boldsymbol\tau_j=1$, so that the
isoscalar-spin operators of the Hamiltonian can be recast in a more
convenient form
\begin{gather}
V=\sum_{i<j}\sum_{p=1}^6 v_p(r_{ij})O^{(p)}(i,j)
V_{SI}+V_{SD}= 
\nonumber \\
=V_{SI}+{1\over2}\sum_{i\alpha,j\beta}\sigma_{i\alpha}
A_{i\alpha,j\beta}^{(\sigma)}\sigma_{j\beta} \,,
\label{eq:v6pott}
\end{gather}
where Latin indices label nucleons, Greek indices label Cartesian
components, and
\begin{equation}
V_{SI}=\sum_{i<j}\left[v_1(r_{ij})+v_2(r_{ij})\right] \,,
\end{equation}
is the spin-isospin independent part of the interaction. The 3A by
3A matrix $A^{(\sigma)}$ contains the interaction between nucleons of
other terms:
\begin{eqnarray}
A_{i\alpha,j\beta}^{(\sigma)}&=&\left[v_3(r_{ij})
+v_4(r_{ij})\right]\delta_{\alpha\beta}+
\nonumber \\
&&\left[v_5(r_{ij})+v_6(r_{ij})\right] (3\hat r_{ij}^\alpha\hat r_{ij}^\beta
-\delta_{\alpha\beta}) \,.
\end{eqnarray}
The matrix $A$ is zero along the diagonal (when $i=j$), in order to avoid
self interaction, and is real and symmetric, with real eigenvalues and
eigenvectors given by
\begin{equation}
\sum_{j\beta}A_{i\alpha,j\beta}^{(\sigma)}\psi_{n,j\beta}^{(\sigma)}=
\lambda_n^{(\sigma)}\psi_{n,i\alpha}^{(\sigma)} \,.
\end{equation}
The matrix $A^{(\sigma)}$ has $n=1...3A$ eigenvalues and eigenvectors.
We can then define a new set of operators written in terms of eigenvectors
of the matrix $A$:
\begin{equation}
\label{eq:oper}
O_n^{(\sigma)}=\sum_{j\beta}\sigma_{j\beta}\psi_{n,j\beta}^{(\sigma)} \,.
\end{equation}
The spin-dependent part of Eq. \ref{eq:v6pott} becomes
\begin{equation}
V_{SD}={1\over2}\sum_{n=1}^{3A} O_n^{(\sigma)2}\lambda_n^{(\sigma)} \,,
\end{equation}
and the corresponding propagator is then
\begin{equation}
e^{-\frac{1}{2}\sum_n O_n^{(\sigma)2}\lambda_n^{(\sigma)}\Delta\tau} \,.
\end{equation}
At first order in $\Delta\tau$ this is equivalent to
\begin{equation}
\prod_n e^{-\frac{1}{2}O_n^{(\sigma)2}\lambda_n^{(\sigma)}\Delta\tau} \,.
\end{equation}
Each factor can be linearized with respect to the operators $O$ by using
the Hubbard-Stratonovich transformation
\begin{equation}
\label{eq:hs}
e^{-\frac{1}{2}\lambda \hat O^2}=
\frac{1}{\sqrt{2\pi}}\int_{-\infty}^\infty dx 
e^{-\frac{x^2}{2}+\sqrt{-\lambda}x\hat O} \,.
\end{equation}
The Green's function then becomes:
\begin{eqnarray}
\label{eq:gf}
G(\bold R,\bold R',\Delta\tau)=
\Big({m\over2\pi\hbar^2\Delta\tau}\Big)^{3A\over2}e^{-{m|\bold R-\bold R'|^2
\over2\hbar^2\Delta\tau}} e^{-V_{SI}(\bold R)\Delta\tau}
\nonumber \\
\times\prod_{n=1}^{3A}{1\over\sqrt{2\pi}}\int dx_ne^{-{x_n^2\over2}}
e^{\sqrt{-\lambda_n\Delta\tau}x_nO_n} \,.
\end{eqnarray}
The newly introduced variables $x_n$, called \emph{auxiliary fields},
are sampled in order to evaluate the integral of Eq. \ref{eq:gf}.
For a spin state given by a product of single particle spin functions,
the linearized Green's function has the effect of changing the spin
state by independently rotating the spin of each single nucleon.

The sampling of auxiliary fields to perform the integral of
Eq. \ref{eq:hs} eventually gives the same effect as the propagator with
quadratic spin operators acting on a trial wave function containing all
the possible good spin states.  The effect of the Hubbard-Stratonovich
is then to reduce the dependence of the number of operations needed
to evaluate the trial wave function from exponential to linear in the
number of nucleons.  The price to pay is the additional computational
cost due to the diagonalization of $A$ matrices and the sampling of the
integral over auxiliary fields.

Sampling of auxiliary fields can be achieved in several ways. The most
intuitive one, in the spirit of Monte Carlo sampling is to consider the
Gaussian in the integral Eq. \ref{eq:hs} as a probability distribution.
The sampled values are then used to determine the action of the operators
on the spin part of the wave function. This is done exactly as in the
diffusion process.  It is possible to use other techniques to evaluate the
integral (e.g. with the three-point Gaussian quadrature\cite{koonin97})
but results must be equivalent after the integration.

The method used to include the spin-orbit and three-nucleon
interaction in the propagator,
and a detailed description of the AFDMC method can be found in
Refs. \cite{sarsa03,pederiva04,gandolfi07c}.

\subsection{Importance sampling}
\label{subsec:sampling}

In order to reduce the variance of estimators, the importance sampling
is required.  In practice a diffusion Monte Carlo
calculation is performed using a Green's
function modified as follows:
\begin{equation}
\label{eq:is}
\tilde G(\bold R,\bold R',\Delta\tau)=
\frac{\psi_I(\bold R')}{\psi_I(\bold R)}G(\bold R,\bold R',\Delta\tau) \,.
\end{equation}

The so called \emph{importance function} $\psi_I$ in the above equation
is often the same as that used for the projection of the energy, and is
evaluated at the walker configuration. More precisely we define
\begin{equation}
\psi_I(\bold R)=\langle\psi_I\vert \bold R\rangle \,.
\end{equation}
     
In this case the distribution function that is sampled in the imaginary
time converges to the quantity
\begin{equation}
f(\bold R,\tau\rightarrow\infty)=\psi_I(\bold R)\phi_0(\bold R) \,.
\end{equation}

The propagator becomes a shifted Gaussian with a modified weight
\begin{gather}
G_0(\bold R,\bold R^\prime,\Delta\tau)=
\left(\dfrac{1}{2\pi D\Delta\tau}\right)^\frac{3A}{2}
e^{-\frac{\left|\bold R-\bold R^\prime
+D\Delta\tau\frac{\nabla\psi_I(\bold R)}{\psi_I(\bold R)}\right|^2}
{2D\Delta\tau}} \,,
\nonumber \\
w=e^{-\left(\frac{E_L(\bold R)+E_L(\bold R^\prime)}{2}-E_T\right)\Delta\tau} \,,
\end{gather}
where $D=\hbar^2/m$ is the diffusion constant, and
\begin{equation}
E_L(\bold R)=-\frac{\hbar^2}{2m}\frac{\nabla^2\psi_I(\bold R)}{\psi_I(\bold R)}+
\frac{V(\bold R)\psi_I(\bold R)}{\psi_I(\bold R)}
\end{equation}
is the local energy of the system.  The additional term in the Gaussian
is often called \emph{drift}, so each walker's configuration is diffused
according to
\begin{equation}
\label{eq:drifted}
\bold R^\prime=\bold R+D\Delta\tau d+\boldsymbol\eta \,,
\end{equation}
where the quantity 
\begin{equation}
d=\frac{\nabla\psi_I(\bold R)}{\psi_I(\bold R)}
\end{equation}
is the drift term, and $\bold\eta$ is a Gaussian random vector.

Importance sampling can also be included in the Hubbard-Stratonovich
transformations that rotate nucleon spinors.  For auxiliary fields
importance sampling is achieved by ``guiding'' the rotation given by each
$O_n$ operator.  More precisely one can consider the following identity:
\begin{align}
&-\frac{x_n^2}{2}+\sqrt{-\lambda_n\Delta\tau}x_nO_n=
\nonumber \\
&-\frac{x_n^2}{2}+\sqrt{-\lambda_n\Delta\tau}x_n\langle O_n\rangle
+\sqrt{-\lambda_n\Delta\tau}x_n\left(O_n-\langle O_n\rangle\right) \,,
\nonumber
\end{align}
where the mixed expectation value of the operator (see the next subsection
for details) is evaluated in the old spin configuration:
\begin{equation}
\langle O_n\rangle=
\frac{\langle\psi_I\lvert O_n\rvert R,S\rangle}
{\langle\psi_I\vert R,S\rangle} \,.
\end{equation}
This can be implemented by shifting the Gaussian used to sample auxiliary
fields, and considering the extra terms in the weight for branching
\begin{align}
\label{eq:hsis}
&e^{-{x_n^2}/{2}+\sqrt{-\lambda_n\Delta\tau}x_nO_n}=
\nonumber \\
&e^{-{(x_n-\bar x_n)^2}/{2}}e^{\sqrt{-\lambda_n\Delta\tau}x_nO_n}
e^{\bar x_nx_n-{\bar x_n^2}/{2}} \,,
\end{align}
where
\begin{equation}
\bar x=\sqrt{-\lambda_n\Delta\tau}\langle O_n\rangle \,.
\end{equation}
The additional weight term in Eq. \ref{eq:hsis} can also be included as 
a local potential, so it becomes
\begin{equation}
e^{-\dfrac{\langle\psi_I\lvert V\rvert R,S\rangle}
{\langle\psi_I\vert R,S\rangle}\Delta\tau} \,.
\end{equation}
By combining the diffusion, the rotation and all the additional factors it is possible to write an explicit propagator
\begin{align}
G(\bold R,\bold R',\Delta\tau)&=G_0(\bold R,\bold R^\prime,\Delta\tau)
\nonumber \\
\times
&e^{ -\left(-\frac{\hbar^2}{2m}
\frac{\nabla^2 |\psi_I (R,S)|}{|\psi_I(R,S)|}
+ \frac{\langle \psi_I|V|RS\rangle}{\langle\psi_I|RS \rangle } -E_0
\right ) \Delta t}
\nonumber \\
\times
&\dfrac{\psi_I(R^\prime,S)}{\psi_I(R,S)}
\frac{|\psi_I(R,S)|}{|\psi_I(R^\prime,S)|}  \,,
\end{align}
where in the choice of the drift term is
\begin{equation}
d=\frac{\nabla|\psi_I(R,S)|}{|\psi_I(R,S)|} \,.
\end{equation}

\subsection{Computation of expectation values}
\label{sec:mix}

The projected walker distribution obtained with the AFDMC is used to
compute expectation values.  For a generic operator $O$ in the limit
$\tau\rightarrow\infty$ the ``mixed'' expectation value is computed as
\begin{equation}
\langle O\rangle_{mix}=\dfrac{\langle\phi_0(\bold R)\vert O
\vert\psi_T(\bold R)\rangle}
{\langle\phi_0(\bold R)\vert\psi_T(\bold R)\rangle} \,.
\end{equation}
We are interested in the expectation value over the ground state $\phi_0$.
Assuming that $\psi_T$ is a good approximation of the ground-state a
better estimate of the ground-state expectation value can be obtained
by combining the variational Monte Carlo and the diffusion
Monte Carlo estimators
in this way:
\begin{equation}
\label{eq:omix}
\langle O\rangle=2\langle O\rangle_{mix}-\langle O\rangle_v \,,
\end{equation}
where $\langle O\rangle_v$ is the expectation value computed over the
variational wave function $\psi_T$ used as trial wave function.

The evaluation of the energy of the system is a particular case, and
can be directly calculated from the projected distribution.  Since the
propagator commutes with the Hamiltonian (but this will change in the
next section when we introduce a constraint), we have
\begin{align}
&\langle H\rangle_{mix}=
\frac{\langle\phi_0(\bold R)\vert H\vert\psi_T(\bold R)\rangle}
{\langle\phi_0(\bold R)\vert\psi_T(\bold R)\rangle}=
\frac{\langle\psi_T(\bold R)\vert H\vert\phi_0(\bold R)\rangle}
{\langle\psi_T(\bold R)\vert \phi_0(\bold R)\rangle}=E_0
\end{align}
The total energy is already the correct value and
since it does not contain a linear error from the trial function, it
does not require the extrapolation of Eq. \ref{eq:omix}.

The propagator used in our AFDMC calculations is written to include
only the first eight operators of the Argonne interactions.  However,
in some case, we can also evaluate the expectation value of the full
Argonne $v_{18}$
Hamiltonian.  In light nuclei, the expectation value of
Argonne $v_8^\prime$
is within few percent of Argonne $v_{18}$\cite{pieper01}. It
is then reasonable to propagate the wave function using the Argonne
$v_8^\prime$
and evaluate the difference between Argonne $v_8^\prime$ and $v_{18}$
using the
extrapolation of Eq. \ref{eq:omix}. This procedure was verified in GFMC
calculations\cite{pudliner97}, and we employed this technique in the case
of low density where Argonne $v_8^\prime$ is a very good approximation to
Argonne $v_{18}$.

More precisely, we evaluate the energy using Argonne $v_8^\prime$ in the
propagator (in addition to the three-nucleon interaction),
and we add to the total energy the
value of $\langle v_{18}-v_8^\prime\rangle$ evaluated as in \ref{eq:omix}.
We expect this approximation to be accurate if this difference is
small as in light nuclei.

\subsection{Constrained path and fixed-phase approximation}
\label{sec:CP-FP}
As described in the above sections, diffusion Monte Carlo
projects out the ground-state of
a given Hamiltonian in terms of the distribution of the walkers.  However
the density of walkers must always be positive definite\cite{reynolds82}.
For walkers with positive weights,
this condition restricts, in principle, the use of the method to that
class of problems where the trial wave function is always positive or
is node-less, such as for a Bose system in the ground-state.
Algorithms which allow negative weights, such as transient
estimation\cite{schmidt84} generally have exponentially increasing variance.

One way to deal with Fermionic systems is to set artificial boundary
conditions between the positive and negative regions of the trial
wave function. It is possible to define a nodal surface where the
trial wave function is zero and during the diffusion process a walker
that crosses the nodal surface is dropped; this is the fixed-node
approximation\cite{schmidt84,anderson76}, and its application in the
diffusion Monte Carlo
algorithm always gives an upperbound to the true Fermionic ground-state
energy.

In the case of nuclear Hamiltonians or for problems where the trial wave
function must be complex, a constrained-path\cite{zhang03,zhang95,zhang97}
approximation is usually applied to avoid the Fermion sign or phase
problem.  The constrained-path method was originally proposed by
Zhang et al. as a generalization of the fixed-node approximation to
complex wave functions.  In constrained-path,
walkers are constrained to regions where
the real part of the overlap with the trial wave function is positive.
This constrained-path approximation was the original method used
to control the phase problem in the AFDMC algorithm\cite{schmidt99}.
More precisely, we have to consider that even for a complex wave function
drift term for the coordinates must be real.  In the case
of the constrained-path approximation, a natural choice for the drift is
\begin{equation}
d=\frac{\nabla Re[\psi_I(\bold R)]}{Re[\psi_I(\bold R)]} \,.
\end{equation}
Moreover, in order to eliminate the decay of the signal-to-noise ratio it
is possible to impose the constrained-path approximation, by requiring
that the real part of the overlap of each walker with the trial wave
function keeps the same sign.  Thus, one can impose
\begin{equation}
\frac{Re[\psi_I(\bold R^\prime)]}{Re[\psi_I(\bold R)]}>0 \,,
\end{equation}
where $\bold R$ and $\bold R^\prime$ denote the coordinates of the
system after and before the diffusion of a time-step.  If this
condition is violated, the walker is dropped.  This form was
found to give better results and was employed in the past AFDMC
calculations\cite{gandolfi06,gandolfi08}.

An alternative way to control the sign problem is the fixed-phase
approximation.  This method was originally proposed by Carlson for
nuclear systems\cite{carlson87}, and also employed for systems whose
Hamiltonian contains a magnetic field\cite{ortiz93}.

We start with the same condition of the reality of the drift, and we
consider the following expression
\begin{equation}
d=\frac{\nabla\lvert\psi_I(\bold R)\rvert}{\lvert\psi_I(\bold R)\rvert} \,.
\end{equation}
With this choice the weight for branching becomes
\begin{eqnarray}
\label{eq:weight2}
\exp\left[{-\left(-\frac{\hbar^2}{2m}
\frac{\nabla^2\lvert\psi_I(\bold R)\rvert}{\lvert\psi_I(\bold R)\rvert}
+\frac{V\psi_I(\bold R)}{\psi_I(\bold R)}
\right)\Delta\tau}\right]
\nonumber \\
\times\frac{\vert\psi_I(\bold R)\vert}{\vert\psi_I(\bold R')\vert}
\frac{\psi_I(\bold R')}{\psi_I(\bold R)} \,.
\end{eqnarray}
Note that in the above expression there is the usual importance sampling
factor as in Eq. \ref{eq:is}, and an additional factor that corrects
for the particular choice of the drift.

A generic complex wave function can be written as
\begin{equation}
\psi(\bold R)=\vert\psi(\bold R)\vert e^{i\phi(\bold R)} \,,
\end{equation}
where $\phi(\bold R)$ is the phase of $\psi(\bold R)$; the factor
appearing in Eq. \ref{eq:weight2} can be rewritten as
\begin{equation}
\frac{\vert\psi_I(\bold R)\vert}{\vert\psi_I(\bold R')\vert}
\frac{\psi_I(\bold R')}{\psi_I(\bold R)}
=e^{i[\phi(\bold R')-\phi(\bold R)]} \,.
\end{equation}

The fixed-phase approximation constrains the walkers to have the same
phase as the importance function $\psi_I$.  It can be applied by keeping
the real part of the last expression.  To keep fixed the normalization
of the Green's function one has an additional factor in the Green's
function that must be included in the weight:
\begin{equation}
e^{-\frac{\hbar^2}{2m}\left(\nabla\phi\right)^2\Delta\tau} \,.
\end{equation}
This can be automatically included by keeping the real part of the
kinetic energy. In fact:
\begin{equation}
Re\left[\frac{\nabla^2\psi_I(\bold R)}{\psi_I(\bold R)}\right]
=\frac{\nabla^2\lvert\psi_I(\bold R)\rvert}{\lvert\psi_I(\bold R)\rvert}-
\left(\nabla\phi(\bold R)\right)^2 \,.
\end{equation}
The real part of the kinetic energy includes the additional weight term
given by the fixed-phase approximation.

A different derivation to introduce the fixed-phase approximation is
the following. Let us consider the evolution of a complex trial wave
function including the importance sampling:
\begin{equation}
\psi_I^*(\bold R)\psi(\bold R,\tau)=
\int G(\bold R,\bold R',\tau)\psi_I^*(\bold R)\psi(\bold R',0)d\bold R' \,.
\end{equation}
The quantity $\psi_I^\star(\bold R)\psi(\bold R,\tau)$ is not real and
positive definite as required, but it is possible to obtain another
positive density as
\begin{gather}
\vert\psi_I(\bold R)\vert\vert\psi(\bold R,\tau)\vert=
\int G(\bold R,\bold R^\prime,\tau)
\dfrac{\vert\psi_I(\bold R)\vert}{\vert\psi_I(\bold R^\prime)\vert}
e^{i[\phi(\bold R^\prime)-\phi(\bold R)]} 
\nonumber\\
\times\vert\psi_I(\bold R^\prime)\vert\vert\psi(\bold R^\prime,0)
\vert d\bold R^\prime \,.
\end{gather}
In this way we impose that the phase of the trial wave function is the
same of that of $\psi_I$.

Both the constrained-path and the fixed-phase are approximations to deal
with the Fermion sign problem and in principle they should be equivalent
if the importance function is close to the correct ground-state of
the system.

It is important to note that Carlson et al.\cite{carlson99} showed that
within the constrained-path approximation the algorithm does not
necessarily give an upper bound in the calculation of energy. This was
also observed by Wiringa et al. in some nuclear simulations using the
GFMC technique\cite{wiringa00}.  It is not guaranteed that our fixed-phase
calculations give an upperbound.  However, in diffusion Monte Carlo
calculations of the
ground-state of quantum dots, where both a real or a complex trial wave
function can be implemented, the fixed-phase approximation gives a higher
energy that the fixed-node approximation\cite{colletti02}.

\subsection{Trial wave function}
\label{sec:psit}

The trial wave function used as the importance and projection function
for the AFDMC algorithm has the following form:
\begin{equation} 
\label{eq:psi_T}
\psi_I(\bold R,S) =  F_J(\bold R) D(\bold R,S) \,,
\end{equation}
where $\bold R\equiv (\bold r_1,\dots,\bold r_N) $ represent the spatial
and $S\equiv (s_1,\dots ,s_N)$ are the spin states of the system.
The spin assignments $s_i$ consist of giving the two-spinor components
for each neutron, namely
\begin{equation}
\vert s_i\rangle=a_i\lvert\uparrow\rangle+b_i\lvert\downarrow\rangle \,,
\end{equation}
where $a_i$ and $b_i$ are complex numbers, and the
$\{\lvert\uparrow\rangle,\lvert\downarrow\rangle\}$ is the neutron-up
and neutron-down base.

The Jastrow correlation function $F_J(\bold R)$ is symmetric under
the exchange of two particles, and independent of spin.  Its role is to
include the short-range pair correlations in the trial wave function. The
generic form for the Jastrow is
\begin{equation}
F_J(\bold R)=\prod_{i<j}f(r_{ij}) \,,
\end{equation}
where the function $f(r)$ is the solution of a Schr\"odinger-like equation
for $f(r<d)$,
\begin{equation}
-\frac{\hbar^2}{m}\nabla^2f(r)+v(r)f(r)=\lambda f(r) \,,
\end{equation}
where $v(r)$ is the spin-independent part of the nucleon-nucleon
interaction, the
healing distance $d<L/2$ is a variational parameter and $L$ is the size
of the box. For distances $r\ge d$ we impose $f(r)=1$.  The Jastrow
part of the trial wave function in the AFDMC case has only the role of
reducing the overlap of neutrons, therefore reducing the energy variance.
Since it does not change the phase of the wave function, it does not
influence the computed energy value in projections methods. In all the
reported results we then fixed $d=2$ fm or $d=L/2$ if $L/2< 2$ fm.

The antisymmetric part of the trial wave function is usually given by
the ground-state of the non-interacting Fermions, which is written as
a Slater determinant

\begin{equation} 
D(\bold R,S) ={\cal A} \left[\prod_{i=1}^N \phi_\alpha(\bold r_i,s_i)\right]=
{\rm Det}\left\{\phi_\alpha(\bold r_i,s_i)\right\} \,,
\end{equation}
where $\alpha$ is the set of quantum numbers of single-particle orbitals,
and ${\cal A}$ is the antisymmetrization operator.

For neutron matter calculations we choose the antisymmetric part as
the ground state of the Fermi gas, built from a set of plane waves. The
infinite uniform system is simulated with $N$ nucleons in a cubic periodic
box of volume $L^3$. The momentum vectors in this box are
\begin{equation}
\bold k_\alpha=\frac{2\pi}{L}(n_{\alpha x},n_{\alpha y},n_{\alpha z}) \,,
\end{equation}
where $\alpha$ labels the quantum state and $n_x$, $n_y$ and $n_z$ are
integer numbers describing the state.  The single-particle orbitals are
given by
\begin{equation}
\phi_\alpha(\bold r_i,s_i)=e^{i\bold k_\alpha\cdot\bold r_i}
\langle\chi_{s,m_s,\alpha}\vert s_i\rangle
\end{equation}

\subsection{Twist-averaged boundary conditions}
\label{subsec:TABC}

Aside from the effect of the phase of the importance function employed
during the projection in imaginary time, the dependence of the energy on
the number of neutrons is the largest systematic error.
Usually one uses periodic boundary
conditions to reduce finite size effects, and simulations are carried
out by using a number of neutrons filling closed shells of plane waves.
There are still sizable errors in the kinetic energy coming from
the shell structure even at the closed shell filling in momentum
space (1, 7, 19, 27, 33, 57, ...).
In order to establish the effect of the finite size of the system due
to the kinetic energy we imposed twist-averaged boundary conditions\cite{lin01}
on the trial wave function. Within periodic boundary conditions,
the phase, picked up by the wave function as a particle makes a circuit
across the unit cell, can be chosen arbitrarily.
These more general boundary conditions for a wave function are
\begin{equation} 
\label{cond:TABC}
\psi(\bold{r}_1+L\hat{\bold{x}},\bold{r}_2,\ldots)
 = e^{i\theta_x}\psi(\bold{r}_1,\bold{r}_2,\ldots) \,,
\end{equation}
where $L$ is the side of the simulation cell.  The boundary condition
$\theta=0$ gives the usual periodic boundary conditions, and the
more general condition with $\theta\ne 0$, twisted boundary conditions.
If the twist angle is integrated over, the single-particle finite-size
effects, arising from shell effects in filling the plane wave orbitals,
are substantially reduced.
Integrating over twists averages over the
volume of $\bold{k}$ space occupied by the first $N$ Brillouin zones
of the simulation cell. The occupied region is a convex polyhedron
that tends to the Fermi surface in the limit of infinite system size
and has the correct volume at all system sizes.  The twist
averaged
kinetic energy must approach the exact energy always from above since
the single-particle kinetic energy is a convex function of $\bold{k}$.

The integration over angles can be achieved in different ways,
either by modifying the trial wave function during the simulation or by
performing several simulations using different wave functions\cite{lin01}.
In practice, once the density of the system is fixed, we consider a grid
of different ${\bold k}_i$--vectors
\begin{equation} 
\label{TABC:grid}
{\bold k}_{\alpha,i}=\left(2\pi{\bold n}_\alpha+\theta_i\right)/L \,
\end{equation}
within the radius corresponding to the Fermi energy, and for each twist
angle $\theta_i$ a simulation is performed.  The total energy is the
given by averaging all the energy obtained for each wave function.

\subsection{The Algorithm}
\label{subsec:algorithm}

The structure of the AFDMC algorithm consists in the following procedures:
\begin{enumerate}
\item Sample the positions and spins, to give $|R,S \rangle$ for
the  initial walkers, from $|\langle \Psi_I |R,S\rangle|^2$
using Metropolis Monte Carlo.
\item Propagate the spatial degrees of freedom as in the usual diffusion
Monte Carlo with a drifted Gaussian for a time step. That is, each walker
configuration is diffused according to Eq. \ref{eq:drifted}.
\item For each walker, build and diagonalize the potential matrix
$A^{(\sigma)}$.
\item Loop over the eigenvectors, sampling the corresponding shifted
Hubbard-Stratonovich variable and update the spinors for a time step.
Introduce approximate importance sampling of the Hubbard-Stratonovich
variables, as discussed in the previous subsections.
\item Propagate with the spin--orbit interaction, using importance
sampling.
\item Evaluate the real part of the local energy to constrain each walker
to have the fixed-phase as described above. This quantity is also stored
with the corresponding weight to calculate the averaged mixed energy.
\item Iterate from 2 to 6 as long as necessary until convergence of the
energy is reached.
\end{enumerate}

To evaluate the error bars, block averages are calculated and the results
combined over different block sizes until the blocks become uncorrelated
and the error bars independent of block size within statistics.

\section{Results}
\label{sec:results}
\subsection{Test of the fixed-phase approximation}
The AFDMC algorithm combined with the constrained-path approximation
was previously employed by Sarsa et al. to study the neutron matter
equation of state at zero temperature\cite{sarsa03}. In that paper the
Hamiltonian contained both a realistic Argonne $v_8^\prime$ two-nucleon
and the Urbana IX
three-nucleon
interactions; this Hamiltonian is often used to calculate properties
of both symmetric nuclear matter and pure neutron matter.

The constrained-path AFDMC
proved to give very satisfactory results for neutron matter
calculations
with a two- and three-nucleon interactions, but some problems were encountered
in the evaluation of the spin-orbit contribution.  The inclusion of
spin-backflow correlations reduced the discrepancies.  A detailed study
considering a pure nucleon-nucleon
interaction emphasized the problem of constrained-path AFDMC
in dealing with the spin-orbit interaction\cite{baldo04}.  A similar
behavior was found by comparing the constrained-path
AFDMC with the GFMC evaluation
for the energy of 14 neutrons in a periodic box\cite{carlson03}. When
using the same Hamiltonian with the same box truncation used in GFMC
calculations of Ref. \cite{carlson03}, the constrained-path
AFDMC overestimated the
energy of 14 neutrons with an Argonne $v_8^\prime$ interaction.

The AFDMC with the fixed-phase approximation overcomes the
discrepancies previously observed in the estimates of the spin-orbit
contribution to the total energy, as shown in table \ref{tab:PNMbench}.
Without tail corrections the constrained-path
AFDMC energy of 14 neutrons at
$\rho=$0.16 fm$^{-3}$ is 20.32(6) MeV compared to 17.00(27) MeV given by
unconstrained GFMC\cite{carlson03}, while the fixed-phase AFDMC energy
is 17.67(5) MeV, within 3\%, and in much better agreement with unconstrained
GFMC.

\begin{table}[h]
\begin{footnotesize}
\begin{center}
\begin{tabular}{|c|cccc|}
\hline
$\rho$ [fm$^{-3}$] & FP-AFDMC  & CP-AFDMC & CP-GFMC   & UC-GFMC \\
\hline
0.04   & 6.75(7)   &          &  6.43(01) &  6.32(03) \\
0.08   & 10.29(1)  &          & 10.02(02) & 9.591(06) \\
0.16   & 17.67(5)  & 20.32(6) & 18.54(04) & 17.00(27) \\
0.24   & 27.7(5)   &          & 30.04(04) & 28.35(50) \\
\hline
\end{tabular}
\end{center}
\end{footnotesize}
\caption{Fixed phase (FP-AFDMC) energies per particle of 14 neutrons interacting
with the
Argonne $v_8^\prime$ interaction in a periodic box without the
inclusion of finite size effects at various densities. The constrained-path
(CP-AFDMC) of
Ref. \cite{sarsa03}, the constrained-path (CP-GFMC)
and the unconstrained (UC-GFMC) GFMC of Ref. \cite{carlson03}
are also reported for a comparison.  All the energies are expressed
in MeV.}
\label{tab:PNMbench}
\end{table}

For higher densities reported in table \ref{tab:PNMbench} it should be
noted that the constrained-path
GFMC significantly differs from unconstrained GFMC because the
Fermion sign problem becomes more severe and the unconstrained energy
estimation has larger fluctuations.  The convergence can be hard to reach
because the imaginary time evolution of the energy can be carried out
only for very small steps. These reasons could introduce some spurious
effects limiting the accuracy of GFMC for the neutron matter calculation
to densities below 0.08 fm$^{-3}$ \cite{carlson03}.

Preliminary results for the ground-state calculation of neutron drops
by means of the fixed-phase AFDMC show that the spin-orbit contribution is
now in agreement with the GFMC results\cite{gandolfi07c}.  Using
the same Hamiltonian, previous constrained-path AFDMC calculations predicted a
spin-orbit splitting (SOS) in $^7n$ neutron-drop about a half the GFMC
result\cite{pederiva04}.  Instead the fixed-phase
AFDMC estimate is in excellent
agreement with the GFMC one\cite{gandolfi07c}, also for neutron drops
containing up to 13 neutrons\cite{gandolfi08c}.

The improvement yielded by using the fixed-phase approximation rather than
the constrained-path is also evident in the comparison of the fixed-phase
AFDMC
with the available constrained-path
AFDMC using spin-backflow correlations.  In table
\ref{tab:AFDMC-PNM} we report all the available calculations computed
within the constrained-path
approximation compared to the fixed-phase one. The corrections
included are only due to the truncation of the nucleon-nucleon
interaction as in the
old calculations\cite{sarsa03}.

\begin{table}[ht]
\begin{footnotesize}
\begin{center}
\begin{tabular}{|c|ccc|}
\hline
$\rho$ [fm$^{-3}$] &FP-AFDMC(14)&CP-AFDMC(14)&JSB-AFDMC(14) \\
\hline
0.12   & 14.52(5) &14.80(9) & \\
0.16   & 19.03(7) &19.76(6) & \\
0.20   & 24.49(5) &25.23(8) & \\
0.32   & 46.60(8) &48.4(1)  & 46.8(1) \\ 
\hline
\hline
$\rho$ [fm$^{-3}$] &FP-AFDMC(66)&CP-AFDMC(66)&JSB-AFDMC(66) \\
\hline
 0.12  & 15.04(8)   &15.26(5) & \\  
 0.16  & 20.14(5)   &20.23(9) & \\
 0.20  & 26.21(5)   &27.1(1)  & \\
 0.32  & 52.47(4)   &54.4(6)  & 52.9(2)\\
\hline
\end{tabular}
\end{center}
\end{footnotesize}
\caption{Fixed phase (FP-AFDMC) energies per particle of 14 and 66 neutrons
interacting with the Argonne $v_8^\prime$+Urbana-IX
interaction in a periodic box
at various densities compared with the available constrained-path (CP-AFDMC)
ones of
Ref. \cite{sarsa03}.  The constrained-path AFDMC
results using a Jastrow-Slater-backflow
(JSB) wavefunction\cite{brualla03} are also shown. In order to make the
comparison possible, the finite size effect due to the truncation of
nucleon-nucleon
interaction was included, while that of Urbana-IX was omitted.  All the
energies are expressed in MeV.}
\label{tab:AFDMC-PNM}
\end{table}

\subsection{Equation of state of neutron matter}
\label{EOS:neutron}
We employed the fixed-phase AFDMC method to study neutron matter by simulating
different numbers of neutrons interacting with the Argonne $v_8^\prime$
potential, including finite-size corrections as described in
Ref. \cite{sarsa03}.  All of the fixed-phase AFDMC results are reported in table
\ref{tab:neutmatAV8}, which shows the energy per neutron of neutron matter for
different densities by varying the number of neutrons.

\begin{table}
\begin{center}
\begin{tabular}{|c|ccc|c|}
\hline
$\rho$ [fm$^{-3}$] & E/N (14)  & E/N (38) & E/N (66) \\
\hline
0.12   & 12.08(5)  & 11.18(4)  & 12.65(4)  \\
0.16   & 14.87(9)  & 13.50(5)  & 15.43(3)  \\
0.20   & 17.6(1)   & 16.10(4)  & 18.27(5)  \\
0.24   &           &           & 21.56(5)  \\
0.28   &           &           & 25.05(6)  \\
0.32   & 27.2(1)   & 25.2(1)   & 28.93(7)  \\
0.36   &           &           & 33.05(6)  \\
0.40   &           &           & 37.15(8)  \\
0.48   &           &           & 46.7(1)   \\
0.56   &           &           & 57.64(9)  \\
0.64   &           &           & 69.90(8)  \\
0.80   & 91.5(2)   & 89.2(2)   & 97.4(1)   \\
\hline
\end{tabular}
\end{center}
\caption{Fixed phase AFDMC energies per particle of 14, 38 and 66 neutrons
interacting by the Argonne $v_8^\prime$ potential in a periodic box at various
densities.  The finite-size effects due to the nucleon-nucleon
truncation are included.
All the energies are expressed in MeV.}
\label{tab:neutmatAV8}
\end{table}

Some finite-size effects are present, as can be deduced by observing the
energies for different numbers of neutrons. The same behavior is followed
at each density, and $E(38)<E(14)<E(66)$. This trend directly follows
the kinetic energy oscillations of $N$ free Fermions which for $N=38$
is lower than either $N=14$ or $N=66$.

\begin{table}[ht]
\begin{center}
\begin{tabular}{|c|cccc|c|}
\hline
$\rho$ [fm$^{-3}$] & E/N(14) & E/N(38) & E/N (66)& E/N (114) \\
\hline
0.12   &  14.77(7) & 13.68(3) &  15.18(2) & 16.05(4) \\
0.16   &  19.41(7) & 18.32(4) &  20.04(2) & 21.31(4) \\
0.20   &  25.05(7) & 24.06(4) &  26.13(4) & 27.82(5) \\
0.24   &  31.74(6) &          &  33.64(4) &  \\
0.28   &  39.79(3) &          &  42.51(3) &  \\
0.32   &  48.61(5) & 48.76(6) &  51.84(2) & 55.13(6) \\
0.36   &  60.03(5) &          &  64.89(5) &  \\
0.40   &  72.38(5) &          &  78.59(6) &  \\
0.48   & 102.74(5) &          & 111.69(9) &  \\
0.56   & 139.8(1)  &          & 152.81(2) &  \\
0.64   &           &          & 202.19(9) &  \\
0.80   &           & 320.3(1) & 328.19(6) &  \\
\hline
\end{tabular}
\end{center}
\caption{Fixed phase AFDMC energies per particle of 14, 38, 66 and 114
neutrons interacting by the Argonne $v_8^\prime$+Urbana-IX
potential in a periodic
box at various densities. Note the difference with the values of table
\ref{tab:AFDMC-PNM} due to the different treatment of finite size effects,
that, in this case, include two- and three-nucleon interaction
contributions.  All the energies
are expressed in MeV.}
\label{tab:allPNM}
\end{table}

For neutron matter the Urbana-IX three-nucleon force reduces to a pairwise spin
interaction modulated by the spectator neutron as explained in
Refs. \cite{pederiva04,gandolfi07c} which can be easily included into
the propagator. Finite-size corrections due to the Urbana-IX can be included
in the same way as for the nucleon-nucleon
interaction, although their contribution is
very small compared to the potential energy. Their effect is appreciable
only for a small number of particles and at large density, i.e. if the
size of the simulation box is small.

All the fixed-phase AFDMC results of 14, 38, 66 and 114 neutrons interacting
with the Argonne $v_8^\prime$
Urbana-IX Hamiltonian, and including all the finite
size effects due to the truncation of two- and three-nucleon
interactions are summarized in table
\ref{tab:allPNM}.  Important finite size effects are still present. The
value closest to the thermodynamic limit is that for 66 neutrons, because
the free Fermi gas energy of this particular system is very similar to
that of the infinite one. However the difference of the energy of 66 and
114 neutrons is always within 6--7\%.  This behavior was also observed
in a study of finite-size effects using the periodic box FHNC
technique\cite{fantoni01c}, and the analysis of Sarsa et al.\cite{sarsa03}
suggests that the energy of the system in the infinite limit is somewhere
between the energies of 66 and 114 neutrons.

\begin{figure}
\vspace{0.5cm}
\begin{center}
\includegraphics[width=7.5cm]{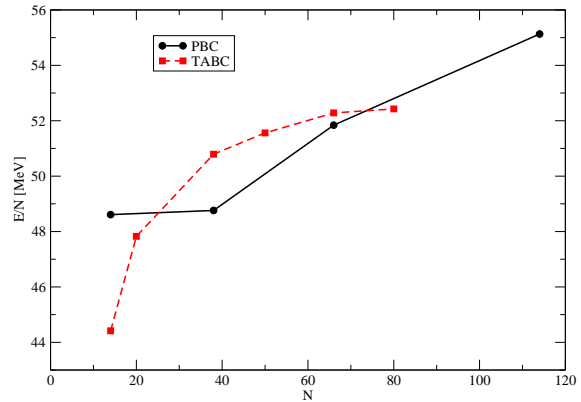}
\vspace{0.5cm}
\caption{(color online) Convergence of the computed energy at
$\rho$=0.32 fm$^{-3}$ as a function of neutrons in a box within the
grid twist-averaging method (TABC) described in the text with ten twists: the
Argonne $v_8^\prime$+Urbana-IX
Hamiltonian were considered.  The equation of state
is compared with
the fixed-phase AFDMC calculations with periodic boundary conditions
(PBC) shown by solid lines.}
\label{fig:PNMtwist}
\end{center}
\end{figure}

In order to better understand the finite size effects due to the kinetic
energy, we repeated several simulations by imposing twist-averaged
boundary conditions in the trial
wave function. The results are displayed in Fig. \ref{fig:PNMtwist},
where we reported the energy obtained by averaging all the results using
sets of ten twist angles in each dimension.  The different behavior of
the energy as a function of the number of neutrons using periodic or
twist-averaged boundary conditions
is
well evident. As expected the effect of twist averaging
is to reduce the jumps of
the energy as a function of $N$ given by periodic boundary conditions.
Then the extrapolation
to the infinite limit of $N$ is better evident using twist averaging. However it
is remarkable that the energy of 66 neutrons computed using either
twist averaging
or periodic boundary conditions
is almost the same. This essentially follows the fact that the
kinetic energy of 66 fermions approaches the infinite limit very well.
In addition, the twist averaging
could be very useful to simulate systems for which
an arbitrary number of Fermions is needed\cite{fantoni01}.

\begin{figure}
\vspace{0.5cm}
\begin{center}
\includegraphics[width=7.5cm]{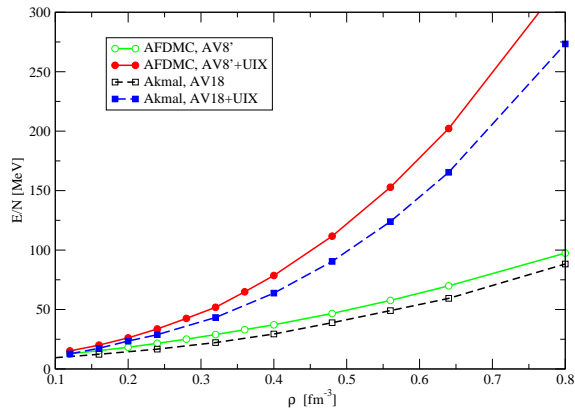}
\caption{(color online) The fixed-phase AFDMC equation of state evaluated
by simulating 66 neutrons in a periodic box; the Argonne
$v_8^\prime$ (AV8$^\prime$)and
Argonne $v_8^\prime$+Urbana-IX (AV8$^\prime$+UIX)
Hamiltonians were considered. The equations of state
are compared with
the variational calculations of Ref. \cite{akmal98} using the Argonne
$v_{18}$ (AV18) and the Argonne $v_{18}$
+Urbana-IX (AV18-UIX) Hamiltonians.  See the legend for details.}
\label{fig:PNMEOS}
\end{center}
\end{figure}

In Fig. \ref{fig:PNMEOS} we plot the fixed-phase AFDMC equation of state,
obtained with the energy of 66 neutrons, and the calculation of Akmal et al. of
Ref. \cite{akmal98}, where the Argonne
$v_{18}$ interaction combined with the
Urbana-IX three-nucleon interaction
was considered.  As it can be seen both the Argonne $v_8^\prime$
and $v_{18}$ give an equation of state
showing essentially the same behavior, with a
difference in the energy that is similar throughout the considered range
of densities.  The addition of the three-nucleon interaction
increases the differences between
the AFDMC and the Akmal et al., in particular at higher densities, implying a
strong difference in pressure and compressibility.

The Argonne $v_8^\prime$
interaction should be more attractive than Argonne $v_{18}$ as shown
in light nuclei and in neutron drop calculation\cite{pudliner97}.
The result shown in Fig. \ref{fig:PNMEOS}, where we compare
Argonne $v_8^\prime$ result with Akmal et al.'s $v_{18}$ values,
do not show this. We believe this is indicative of systematic errors
in the FHNC/SOC calculations.
The fixed-phase AFDMC has proved to be in very good agreement with the
GFMC results for light nuclei\cite{gandolfi07b}, and also with the GFMC
results for 14 neutrons.  On the other hand the fixed-phase AFDMC calculation of
the nuclear matter suggested that the FHNC/SOC approximation could miss
important contributions, in particular those coming from the neglected
elementary diagrams in the FHNC summation\cite{gandolfi07}. In the case
of Akmal et al.
calculations the energy is computed by means of a cluster expansion
for which no evidence of convergence can be provided.  The addition of the
Urbana-IX three-body interaction to the Hamiltonian increases the differences
between the AFDMC results and of Akmal et al.
ones, and, again, this confirms
that the variational technique based on the cluster expansion
gives a lower energy because it neglects important contributions.
However, we stress the fact that in the case of neutron matter the
contribution of the tensor-$\tau$ force is small compared to the other
channels of the interaction. For such reason the calculation of the energy
within traditional variational techniques based on FHNC/SOC or cluster
expansion could be more accurate for pure neutron matter without protons.
This is not true in dealing with nuclear matter where the effect of
tensor-$\tau$ is most important, as confirmed in ref. \cite{gandolfi07}.

The AFDMC results have been fitted with the following functional form:
\begin{equation}
\label{eq:fit}
\frac{E}{N}(\rho)=a\rho^{\beta}+c\rho^{\gamma},
\end{equation}
where $E/N$ is the energy per neutron in MeV as a function of the density
in fm$^{-3}$.  The parameters of the fit for both
Argonne
$v_8^\prime$ and the
full Argonne $v_8^\prime$+Urbana-IX
Hamiltonian are reported in table \ref{tab:fitEOS}.
We also tried to use the functional form of Ref \cite{nicotra06} where
$\beta=1$. We had a worse $\chi^2$ but the equation of state
and the pressure as a
function of the density does not change in a significant way.

\begin{table}[h!]
\begin{center}
\begin{tabular}{|c|cccc|}
\hline
Hamiltonian & a  & c & $\beta$ & $\gamma$ \\
\hline
AV8$^\prime$     & 23.0 & 115.7 & 0.37 & 1.87 \\
AV8$^\prime$+UIX & 32.6 & 507.8 & 0.48 & 2.375 \\
\hline
\end{tabular}
\end{center}
\caption{The parameters of Eq. \ref{eq:fit} fitting the equation of
state computed
with the full Argonne $v_8^\prime$+Urbana-IX (AV8$^\prime$+UIX)
Hamiltonian and with the nucleon-nucleon
interaction
only (AV8$^\prime)$.
The parameters $a$ and $c$ are expressed in [MeV/fm$^{-3}$].}
\label{tab:fitEOS}
\end{table}

\subsection{Argonne $v_8^\prime$ and $v_{18}$ interactions}
As described in the above sections, in most cases the Argonne
$v_{18}$ result
is evaluated as a perturbation of the Argonne
$v_8^\prime$\cite{pudliner97}.
The assumption is reasonable since the Argonne $v_8^\prime$ potential contains
most of the contributions of $v_{18}$ potential
and was obtained with a reprojection by
keeping only the most important terms.  However the operators appearing
in Argonne $v_{18}$ and not in Argonne $v_8^\prime$
are not exactly included in the GFMC
calculations. The imaginary-time GFMC evolution is performed using
Argonne $v_8^\prime$
and the energy is calculated perturbatively in the difference
between $v_8^\prime$ and $v_{18}$, which for nuclei is a fraction of an MeV.

This method is also used in the FHNC/SOC calculation where only the
lowest order two-body nucleon-nucleon
correlations are included in the variational
wave function\cite{akmal98}. However there is no reason to believe that
such a calculation gives an upper bound to the true energy, and this
approximation may not be good, particularly for higher densities.

When using a propagator including the Argonne $v_8^\prime$ potential the
difference between the energies computed using Argonne
$v_8^\prime$ and $v_{18}$ is
actually very small. For instance, for 14 neutrons at $\rho\leq$0.12
fm$^{-3}$ the difference between Argonne $v_8^\prime$ and
$v_{18}$ is less than 2 MeV
per neutron and is 2.7 MeV and 5.1 MeV for densities of 0.16 fm$^{-3}$
and 0.20 fm$^{-3}$ respectively.  On the other hand, a plain truncation
of Argonne $v_{18}$
in the propagator leads to huge energy differences in the two
estimates.  At $\rho$=0.12 fm$^{-3}$ the energy of 14 neutrons with
the Argonne $v_8^\prime$+Urbana-IX Hamiltonian is 14.12 MeV,
while it is 3.60 MeV for
Argonne $v_{18}$+Urbana-IX.
This means that the extra $v_{18}$ terms cannot be thought of as a
small correction to Argonne $v_8^\prime$,
at least in this range of densities.
However, at $\rho\leq$0.04 fm$^{-3}$ the difference between Argonne
$v_8^\prime$
and $v_{18}$ is a few percent of the total energy, so we can safely evaluate
this difference as a perturbation using the $v_8^\prime$ propagator.

In the very low-density regime the neutron matter is a superfluid gas,
and a trial wave function written in a BCS form including explicitly
the pairing between neutrons is needed\cite{gandolfi08b,fabrocini05}.
However, we expect that the expectation value of the Argonne $v_8^\prime$
interaction to be of order of that of Argonne $v_{18}$
both in the superfluid and
in the normal phase.  Here we are only interested to a qualitative study
of the difference between Argonne $v_8^\prime$ and $v_{18}$,
thus a wave function
as presented in section \ref{sec:psit} was used, rather then that of
Ref. \cite{gandolfi08}.

It is interesting to focus on the equation of state
of neutron matter in the
low-density regime and in the normal phase.  The Argonne
$v_{18}$+Urbana-IX Hamiltonian
as described was used. The range of $\rho\le$0.04 fm$^{-3}$ is
particularly relevant in the study of properties of the inner crust
of neutron stars.  The very low density neutron matter approaches a
regime which is almost universal, and is analogous to, for instance,
cold atoms\cite{gezerlis08}.  In this regime many-body techniques
can be compared and calibrated\cite{borasoy08,lee08}.

\begin{table}[ht]
\begin{center}
\begin{tabular}{|l|c|c|}
\hline
$\rho$ [fm$^{-3}$] & $E/N$ & $\langle$AV8$^\prime$-AV18$\rangle$ \\
\hline
$3.377\times10^{-5}$  & 0.089(1) &   -0.00197574  \\
$2.702\times 10^{-4}$ & 0.367(2) &   0.0002776    \\
0.002162              & 1.289(2) &   0.002525     \\
0.007295              & 2.606(4) &   0.021712     \\
0.01729               & 4.277(7) &   0.082534     \\
0.03377               & 6.197(2) &   0.30802      \\
\hline
\end{tabular}
\end{center}
\caption{Fixed phase AFDMC
energies per particle 66 neutrons interacting with the Argonne $v_8^\prime$
+Urbana-IX interaction in a periodic box at various densities. The
difference between the $v_8^\prime$ and the $v_{18}$ interactions
(AV8$^\prime$-AV18),
evaluated with an extrapolation as described in Sec. \ref{sec:mix}
is also reported.  All the energies are expressed in MeV.}
\label{tab:lowPNM}
\end{table}

We report the energy of 66 neutrons in a periodic box in table
\ref{tab:lowPNM}. The Hamiltonian uses the Argonne
$v_8^\prime$+Urbana-IX potential, and the potential
is corrected for finite size effects.  The difference between
the Argonne $v_8^\prime$ and the $v_{18}$ interactions was extrapolated as
described in Sec. \ref{sec:mix}.  As it can be seen, at such densities
the Argonne $v_{18}$ contribution is similar to
$v_8^\prime$, of the order
of few percent with respect to the total energy. We believe that such a
difference would be even smaller if the full Argonne $v_{18}$ was implemented in
the propagator, and then evaluated without the extrapolation described
in the above sections.

\begin{figure}[ht]
\vspace{0.5cm}
\begin{center}
\includegraphics[width=7.5cm]{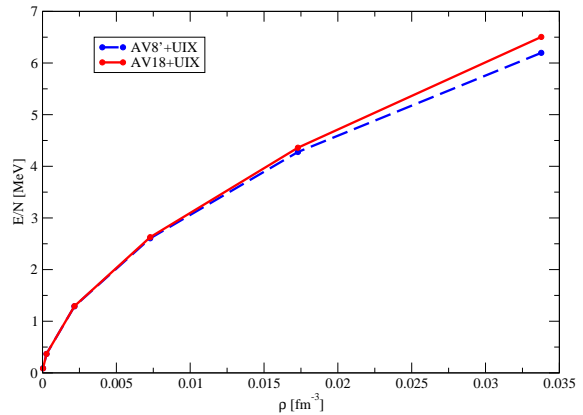}
\caption{(color online) The equation of state
of neutron matter in the low-density
regime. The Argonne $v_8^\prime$ (AV8$^\prime$)
and $v_{18}$ (AV18) interactions combined with the Urbana-IX (UIX)
three-nucleon interaction were considered as indicated in the legend.}
\label{fig:lowEOS}
\end{center}
\end{figure}

The equation of state
of neutron matter in the low-density regime is reported in
Fig. \ref{fig:lowEOS}, where the energy per neutron as a function of the
density is calculated both with the AV8$^\prime$ and with the AV18
nucleon-nucleon
interaction combined with the Urbana-IX three-nucleon interaction.
As it can be seen, the difference
between the two Hamiltonians considered is very small in this regime.
The Argonne $v_8^\prime$ and $v_{18}$
combined with the Urbana-IX essentially give the same
energy, and only small deviations are present when the density increase
above $\approx$0.015 fm$^{-3}$. This result is confirmed by the fact
that in such a regime the neutron-neutron interaction is dominated by the S
channel that in the Argonne $v_8^\prime$ is the same as in
$v_{18}$\cite{pieper02}.
There is a small trend that the energies are sensibly higher at
$\rho\leq$0.015 fm$^{-3}$.  Other many-body calculations
are in general not in agreement and present very different behaviors in
this regime\cite{pethick95}.

\section{Conclusions}
We accurately calculated the equation of state of neutron matter using
the auxiliary field diffusion Monte Carlo method.  We started from a
non-relativistic nuclear Hamiltonian containing two-- and three--nucleon
potentials.  The AFDMC algorithm suffers from the usual fermion sign problem
present in all fermionic Monte Carlo calculations, and we find that the
fixed-phase approximation used to control it to be more accurate than the
previously used constrained-path approximation.  In particular, in this
work we demonstrated that the fixed-phase AFDMC
overcomes the problems encountered
when dealing with the spin-orbit interaction. The Urbana-IX three-body force is
included in the fixed-phase
AFDMC calculation without any perturbative evaluation
because it is naturally included in the Green's function used for the
propagation.  The fixed-phase AFDMC
reveals some problems of the variational
cluster summation (or FHNC/SOC) technique just highlighted in the nuclear
matter calculation with a simple $v_6$-like interaction\cite{gandolfi07}.

We computed the equation of state
of neutron matter using a modern, but still
simplified, nucleon-nucleon
interaction combined with a realistic three-nucleon interaction
in the regime
of interest for predicting the properties of neutron stars, and we found
some deviations with respect to past variational calculations based
on cluster expansion, in particular at high densities.  The difference
between the Argonne $v_8^\prime$+Urbana-IX Hamiltonian
and that containing the more
sophisticated Argonne $v_{18}$ interaction was perturbatively evaluated in
the low density regime, where the equation of state
is useful to constrain properties
of the inner crust of neutron stars.  Our equation of state
can also be useful to
compare the wide range of Skyrme forces used to study the neutron matter.

We are working to include the full Argonne $v_{18}$ interaction in the
two-body part of the Hamiltonian, and we are investigating the effect
of the more complex Illinois three-nucleon forces.  The effect of those
forces in neutron drops and in neutron matter will be a subject of
future work.

\begin{acknowledgments}
We thank J.~Carlson, S. C.~Pieper and R. B.~Wiringa for useful
discussions. This work was supported in part by NSF grants PHY-0456609
and PHY-0757703.  Calculations were partially performed on the HPC
facility ``BEN'' at ECT$^\star$ in Trento under a grant of Supercomputing
Projects, partially on the HPC facility of SISSA/Democritos in Trieste
and partially on the HPC facility "WIGLAF" of the Department of Physics,
University of Trento.  \end{acknowledgments}

\bibliographystyle{apsrev}
\bibliography{neutmat}

\end{document}